\documentclass[final,authoryear,5p]{elsarticle}

\usepackage[hyphens]{url}

\usepackage{graphicx}
\usepackage{amsmath}
\usepackage{booktabs}
\usepackage{caption}
\usepackage{multicol}
\usepackage{multirow}
\usepackage{array}

\usepackage{amssymb}

\usepackage[%
breaklinks=true,%
colorlinks=true,%
pdfauthor={First Author et al.},%
pdftitle={Template for manuscripts in Advances in Space Research}%
]{hyperref}

\usepackage{etoolbox}
\makeatletter
\patchcmd\@combinedblfloats{\box\@outputbox}{\unvbox\@outputbox}{}{\errmessage{\noexpand patch failed}}
\makeatother

\usepackage[hyphens]{url}

\journal{Advances in Space Research}

\begin{document}

\begin{frontmatter}



\title{On the Anthropogenic and Natural Injection of Matter into Earth's Atmosphere}


\author[label1]{Leonard Schulz\corref{cor}}
\ead{l.schulz@tu-bs.de}

\author[label1,label2]{{\rm Karl-Heinz Glassmeier}}
\ead{kh.glassmeier@tu-bs.de}

\cortext[cor]{Corresponding author}
\address[label1]{Institut f{\"u}r Geophysik und extraterrestrische Physik, Technische Universit{\"a}t Braunschweig, 38106 Braunschweig, Germany}
\address[label2]{Max-Planck-Institut f\"ur Sonnensystemforschung, 37077 G\"ottingen, Germany}


\begin{abstract}
Every year, more and more objects are sent to space. While staying in orbit at high altitudes, objects at low altitudes reenter the atmosphere, mostly disintegrating and adding material to the upper atmosphere. The increasing number of countries with space programs, advancing commercialization, and ambitious satellite constellation projects raise concerns about space debris in the future and will continuously increase the mass flux into the atmosphere. In this study, we compare the mass influx of human-made (anthropogenic) objects to the natural mass flux into Earth's atmosphere due to meteoroids, originating from solar system objects like asteroids and comets. The current and near future significance of anthropogenic mass sources is evaluated, considering planned and already partially installed large satellite constellations. Detailed information about the mass, composition, and ablation of natural and anthropogenic material are given, reviewing the relevant literature. Today, anthropogenic material does make up about 2.8\,\% compared to the annual injected mass of natural origin, but future satellite constellations may increase this fraction to nearly 40\,\%. For this case, the anthropogenic injection of several metals prevails the injection by natural sources by far. Additionally, we find that the anthropogenic injection of aerosols into the atmosphere increases disproportionately. All this can have yet unknown effects on Earth's atmosphere and the terrestrial habitat.
\end{abstract}


\end{frontmatter}

\parindent=0.5 cm

\section{Introduction}
Earth's atmosphere is subject to a constant bombardment by various objects from space. Most are of natural origin, i.\,e. meteoroids (in the following, the term meteoroids refers to objects of natural origin without any size limit) from comets, asteroids, and even differentiated bodies. With the exploration of space, anthropogenic objects like spacecraft and rocket bodies in orbit around Earth also enter the atmosphere. Upon reentry, bodies heat up and ablate depending on their physical and chemical properties. This way, matter in form of atoms and aerosols is injected into the atmosphere.   

With the steady growth of spaceflight activities with evermore nations operating space programs and the increase of commercialization, more and more objects are launched into orbit around Earth. This has raised major concerns about space debris \citep{Book_2006_Klinkrad}. As a result, standards have been introduced to minimize the amount of orbital debris \citep{Tec_ISO24113:2019} and space agencies like ESA and NASA have introduced guidelines and requirements, largely accepting those standards \citep[see for example][]{Tec_ESA_SDM, Tec_NASA-STD-8719.14}. A consequence of these guidelines is that payload launched into low Earth orbit (LEO) has to be disposed of within 25 years after end of operation. This is achieved by reentry into the atmosphere. Hence, more and more anthropogenic material is injected into the atmosphere, raising questions about its significance in comparison to the natural injection caused by the ablation of meteoroids, and about possible impacts on the atmosphere itself. 

Several companies have proposed large satellite constellations of hundreds to thousands of small spacecraft in LEO providing global internet and other telecommunication services \citep{Art_ODQN_Liou_2018}. The amount of spacecraft to be launched combined with their limited lifetime will dramatically increase the anthropogenic amount of mass reentering Earth's atmosphere in the future. Thus, the future influx caused by those satellite constellations needs to be considered in more detail.

In this study, we provide a first overview and comparison of the natural and anthropogenic injection of matter into Earth's atmosphere. We focus on the mass influx, its elemental composition, and the resulting ablation products injected into the atmosphere. This is done separately for natural injection (Section \ref{sec:NATMAT}) and anthropogenic injection (Section \ref{sec:ANTMAT}). The necessary information to qualify the mass influx, the elemental composition, and the ablation processes has been acquired from many published studies, partly providing conflicting numbers and information. We try a best effort summary of all the available information.

With this information we provide a review of the natural injection and three different scenarios for the anthropogenic injection. These scenarios include a present day analysis as well as two near-future scenarios taking into account different planned projects for large satellite constellations. This allows to compare the relative contributions of human-made objects and natural objects entering the atmosphere.

\section{Natural injection} \label{sec:NATMAT}
Many meteoroids originating from asteroids, comets, material of planetary origin, interplanetary and even interstellar dust \citep[see for example][]{Art_Jewitt_2000,Art_Plane_2017} enter Earth's atmosphere every day. In this section, we look at the mass, composition and ablation of these meteoroids and estimate the resulting injection with respect to ablation products and the elemental composition.

\subsection{Yearly natural mass influx} \label{sec:NATMASS}
\begin{figure*}[!t]
\begin{center}
\includegraphics[width= 1.7\columnwidth]{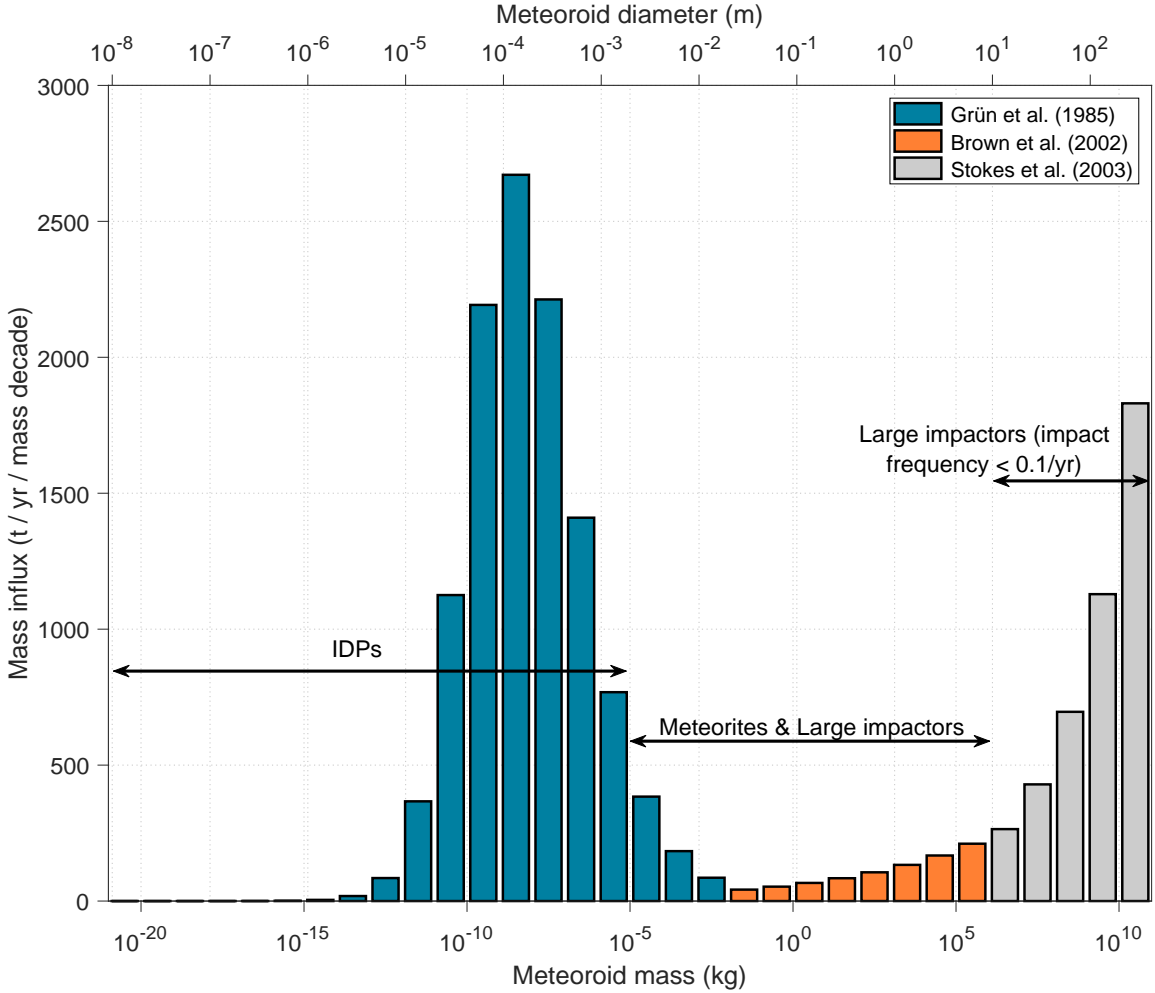}
\end{center}
\caption{Variation of the yearly mass influx distribution of Earth's atmosphere with meteoroid mass. The particle diameter is calculated by using a density of 2500\,kg/m$^3$ and assuming a spherical shape. The blue data, mainly covering the IDP mass range, are based on the flux law of \cite{Art_Gruen_1985}. The orange data, covering the meteorite and large impactor mass range, are calculated using the power law from \cite{Art_Brown_2002} with a mean impact velocity of 20\,km/s. The grey data is calculated using the power law of \cite{Tech_Stokes_2003}, given in \cite{Art_Drolshagen_2017}, again using the same average velocity. It covers large impactors which impact Earth less than every 10 years. The influx distribution reflects the mean model from \cite{Art_Drolshagen_2017} and was calculated as shown in \ref{APP1}. Exact values are given there, too. The annual mass input is 11,509\,t for the \cite{Art_Gruen_1985} model, 863\,t from \cite{Art_Brown_2002}, and 4,349\,t from \cite{Tech_Stokes_2003}.}
\label{fig1}
\end{figure*}

The knowledge of the total natural mass flux into Earth's atmosphere is of high importance. The mass influx distribution is sort of bimodal with a maximum at a particle mass of about 10$^{-8}$\,kg \citep[e.\,g.][]{Book_Flynn_2002, Art_Carrillo_2015, Art_Plane_2017} and a second maximum at high particle masses, although the mass influx increases continuously for large objects. Objects in the size range of a few millimeters to meters, which are the main source of meteorites found on Earth, only contribute a small fraction of the whole mass \citep[][]{Book_Flynn_2002}. The mass influx distribution used in this study is displayed in Figure \ref{fig1}. 

The distribution has two mass ranges with rather high mass input, a small dust particle contribution in the several microns to mm-size range and large meteoroids in the tens of meters range. Differentiation of these ranges is crucial because of the different origin, composition and ablation of those two groups (see Sections \ref{sec:NATCOMP}, \ref{sec:NATABL}). The peak at small meteoroid sizes of several microns to millimeters is caused by the large amount of interplanetary dust particles (IDPs) in the solar system. The particles themselves have very low masses, but are high in number. They have several sources, mainly various types of comets and the asteroid belt, whereas the contribution of interstellar material is negligible \citep{Art_Plane_2017}. 

Dust particles are normally defined to be smaller than tenth of microns \citep{Art_Rubin_2010, Art_Koschny_2017}. However, based on the analysis and modelling of the observations of the zodiacal dust cloud by the Infrared Astronomical Satellite (IRAS) and ground based radars as well as various other observations \citep{Art_Nesvorny_2010, Art_Nesvorny_2011} we adopt a cutoff size of 2\,mm. Therefore, the upper mass limit of the IPD population is roughly 10$^{-5}$\,kg (Fig. \ref{fig1}).

In contrast, the mass flux peak at high impactor sizes is caused by their high mass, while their impact rate is quite low and decreases with increasing size. Bodies heavier than hundreds of tons (larger than several meters in diameter) hit Earth once a year, while impacts with objects several ten meters in diameter occur only once in a thousand years \citep{Art_Chapman_1994, inCol_Zolensky_2006}. However, upon entering Earth's atmosphere, large impactors ablate and disintegrate, leaving behind a trail of aerosols and particles of molecular size. Ablation material in the atmosphere in form of dust particles seems to sediment within several months \citep{Art_Klekociuk_2005, Art_Gorkavyi_2013}, while we can not rule out that ions and particles of molecular size remain for a longer time in the upper atmosphere. Thus, we include large bodies that impact Earth at least every 10 years to account for such ablation and disintegration processes.

\cite{Art_Drolshagen_2017} have calculated the mass influx of meteoroids in a size range of 10$^{-21}$ to 10$^{11}$\,kg. Their mean model for masses below 10$^{-2}$\,kg is based on the widely used interplanetary flux model of \cite{Art_Gruen_1985}, which is used by NASA \citep{Tech_Moorhead_2020} and is close to the newest ESA meteoroid flux model IMEM2 \citep{Art_Soja_2019}. It is derived from different spacecraft in-situ measurements of meteoroids and zodiacal light as well as lunar impact measurements. The model provides an analytic function of the particle flux at 1 AU. For the intermediate mass range, 10$^{-2}$ to 10$^{6}$\,kg, the power law model of \cite{Art_Brown_2002} is used, which is based on spacecraft fireball data. For large bodies heavier than 10$^{6}$, a similar power law adapted from \cite{Tech_Stokes_2003} is used.  

Beside the named studies, \cite{Art_Drolshagen_2017} have included measurements from the Hubble Space Telescope solar array impacts of meteoroids \citep{Misc_McDonnell_2005} as well as visual data from meteor entries \citep{Art_Koschny_Drolshagen_2017} to verify the model of \cite{Art_Gruen_1985}. Additionally, studies from \cite{Art_Halliday_1996} (fireball data) and \cite{Art_Suggs_2014} (lunar impact flashes) were used to find the best way to connect the models of \cite{Art_Gruen_1985} and \cite{Art_Brown_2002}. As \cite{Art_Drolshagen_2017} only briefly reviewed other studies that also provide estimates on the annual mass influx we shortly discuss these other studies.

For the large impactor size range, we regard the models by \cite{Art_Brown_2002} and \cite{Tech_Stokes_2003} as the current best models. However, for the IDP mass range, there is a large difference between the various estimates proposed. \cite{Art_Plane_2012} reviews several studies, experiments, and models regarding the influx of IDPs into Earth's atmosphere. The mass influx estimates vary from 1,800 to 100,000\,t/yr for the respective mass range. Thus, the IDP mass range needs a more critical discussion to derive a suitable estimate for the purpose of our study. Four studies are important here.

\cite{Art_Nesvorny_2010} modelled the zodiacal dust cloud using IRAS data. They estimated an influx of 100,000\,t/yr. A later update \citep{Art_Nesvorny_2011}, using refined orbital characteristics of the IPDs based on meteor radar data, proposes a more realistic value of about 15,000\,t/yr. 

\citet[pp. 148--157]{Book_Hughes_1978} used satellite, radar and visual data of IDP and micrometeor entries to reach an estimate. The IDP mass influx rate of 16,100\,t/yr in a mass range of $10^{-17}$ to $10^{-1}$\,kg is widely accepted and part of the mass influx distribution presented by \cite{Book_Flynn_2002}. 

\cite{Art_Mathews_2001} analyzed observations of micrometeor entries into the upper atmosphere measured by the high power low aperture (HPLA) radar at the Arecibo observatory. They determine the mass and speed of entering particles, reaching estimates of 1,600 and 2,700\,t/yr for the mass range $10^{-8}$ to $10^{-1}$\,kg. These are the lowest annual influx values of all those studies reviewed in \cite{Art_Plane_2012}. Their mean geocentric velocity of 50\,km/s, endorsed by later measurements with Arecibo \citep{Art_Janches_2006}, is much higher than in other observations (\citealp[pp. 150--155]{Book_Hughes_1978}; \citealp{Art_Gruen_1985}; \citealp{Art_Nesvorny_2010, Art_Nesvorny_2011};  \citealp{Art_Koschny_Drolshagen_2017}). Such a large velocity raises questions as it implies that the majority of the dust particles moves retrograde in the solar system. This could be due to HPLA radars being unable to detect very slow ($<$15\,km/s) and small particles. The lack of detection of smaller particles could explain the quite low influx estimate. \cite{inPro_vonZahn_2005} discusses further possible shortcomings and biases. Here, we assume that \cite{Art_Mathews_2001} significantly underestimate the amount of incoming meteors. 

\cite{Art_Brownlee_1993} measured IDP and meteoroid impacts on the Long Duration Exposure Facility (LDEF). Impact crater size, depth and number were determined in order to obtain information about the mass of each impactor. Assumptions had to be made concerning the particle properties, velocity and impact angle. The integration of the derived mass distribution yields approx. 27,000--40,000\,t/yr \citep{Art_Brownlee_1993, Art_Taylor_1998, Art_Mathews_2001}.

The estimate of \cite{Art_Brownlee_1993} has several uncertainties as it is based on a mean geocentric particle velocity of 16.9\,km/s. For further detail on the velocity we refer to \cite{Art_Gruen_1985, Art_Taylor_1995, Art_Taylor_1996, Art_Taylor_Elford_1998, inCol_Brown_2005, Art_Drolshagen_2008, Tech_ECSS_2008, Art_Nesvorny_2010, Art_Nesvorny_2011}, and \cite{Art_Carrillo_2016}. After reviewing all the mentioned literature, we use the higher geocentric mean velocity (normalized on mass) of 20\,km/s for IDPs, as the evaluations of \cite{Art_Taylor_Elford_1998, inCol_Brown_2005} seem to incorporate the best combination of relevance and experimental falsification. A similar velocity is also used by \cite{Art_Drolshagen_2017}. 

Increasing the mean velocity implies a reduction of particle mass. This results in a shift of the mass distribution towards lower masses and reduces the mass influx value. Also considering \cite{Art_Borin_2009, Art_Cremonese_2012}, a realistic mass influx estimate is a value below 20,000\,t/yr.

\subsubsection{Resulting mass influx and mass distribution}
Results of the studies discussed roughly agree with the mass influx estimate by \cite{Art_Drolshagen_2017}. The \cite{Art_Gruen_1985} flux model is still widely accepted, and we use it to determine an annual mass influx for the different mass decades. For the higher mass ranges, the power laws by \cite{Art_Brown_2002} and \cite{Tech_Stokes_2003} are used (Figure \ref{fig1}). For further details see \ref{APP1}.

Thus, considering the mass range of objects impacting Earth less than every 10 years, 10$^{-21}$ to 10$^{6}$\,kg, we yield an annual mass influx of 12,372\,t with most of the mass (11,509\,t) caused by particles with masses lower than 10$^{-2}$\,kg. The mass flux is dominated by the IDP contribution of 10,856\,t/yr. For all the mass ranges studied considerable differences in the estimation of the mass influx exist. With an error factor 2 we yield a range of 6,186 to 24,746\,t/yr for the mass influx.

\subsection{Composition of the natural material} \label{sec:NATCOMP}
IDPs and larger meteoroids have quite different compositions due to different origins. The major contribution to the IDP flux is thought to originate from Jupiter Family Comets (JFCs) (\citealp{Art_Zolensky_2006}; \citealp{Art_Nesvorny_2010}; \citealp[pp. 282--283]{inCol_Jenniskens_2015}; \citealp{Art_Yang_2015}; \citealp{Art_Carrillo_2016}). The composition of IDPs has been examined comprehensively in two studies: \cite{Art_Schramm_1989} examined 200 IDPs on their major element composition and \cite{Art_Arndt_1996} gathered data on 89 IDPs covering elemental abundances also for minor elements. Both studies only give relative abundances, normalized to Si, Fe, and CI carbonaceous chondrite class abundance. The CI abundance is representative of the solar system abundances of elements \citep{Art_Anders_1982, Art_Anders_1989} and used for normalization. The only element with a significant mass fraction not determined by the two above mentioned studies is hydrogen. Here, we use the 2.02\,wt\% value from \cite{Art_Anders_1989}. The absolute elemental mass abundances used in following discussions are listed in Table \ref{TAB1}. For further considerations we use mean values. To summarize, a significant fraction of the mass contribution are metal (31\,\%) and metalloid elements (13\,\%), while the majority is nonmetallic (47\,\%). About 8\,\% of the mass could not be assigned to an element.

\begin{table}[!ht]
\centering
\caption{Elemental composition of IDPs and meteorites.}
\label{TAB1}
\centering
\footnotesize{
\begin{tabular}{@{}rllp{1cm}p{1cm}p{1cm}p{1cm}@{}}
\toprule
Z          & El.      & Unit     & \cite{Art_Arndt_1996} & \cite{Art_Schramm_1989} & Mean IDP & Meteo- rites \\ \midrule
1     & H       & $\mu$g/g &       &         & 20200\textsuperscript{a}    & 186        \\
3     & Li      & $\mu$g/g &       &         &          & 2          \\
4     & Be      & ng/g     &       &         &          & 31         \\
5     & B       & ng/g     &       &         &          & 377        \\
6     & C       & wt\%     &       & 9.7     & 9.7      & 0.3       \\
7     & N       & $\mu$g/g &       &         &          & 69         \\
8     & O       & wt\%     &       & 30.7    & 30.7     & 35.2       \\
9     & F       & $\mu$g/g &       &         &          & 97         \\
11    & Na      & $\mu$g/g & 3747  & 5503    & 4625     & 5936       \\
12    & Mg      & wt\%     & 9.3   & 11.0    & 10.1     & 13.6       \\
13    & Al      & wt\%     & 1.2   & 0.9     & 1.1      & 1.2        \\
14    & Si      & wt\%     & 12.9  &         & 12.9     & 17.1       \\
15    & P       & $\mu$g/g & 1746  &         & 1746     & 1292       \\
16    & S       & wt\%     & 3.6   & 5.3     & 4.4      & 2.1        \\
17    & Cl      & $\mu$g/g & 1229  &         & 1229     & 188        \\
19    & K       & $\mu$g/g & 540   &         & 540      & 764        \\
20    & Ca      & wt\%     & 0.3   & 1.0     & 0.6      & 1.3        \\
21    & Sc      & $\mu$g/g & 12    &         & 12       & 8          \\
22    & Ti      & $\mu$g/g & 549   &         & 549      & 676        \\
23    & V       & $\mu$g/g & 74    &         & 74       & 71         \\
24    & Cr      & $\mu$g/g & 2199  & 3590    & 2894     & 3466       \\
25    & Mn      & $\mu$g/g & 1644  &         & 1644     & 2355       \\
26    & Fe      & wt\%     & 17.3  & 17.9    & 17.6     & 25.9       \\
27    & Co      & $\mu$g/g & 337   &         & 337      & 808        \\
28    & Ni      & wt\%     & 0.5   & 0.7     & 0.6      & 1.7        \\
29    & Cu      & $\mu$g/g & 186   &         & 186      & 91         \\
30    & Zn      & $\mu$g/g & 405   &         & 405      & 57         \\
31    & Ga      & $\mu$g/g & 18    &         & 18       & 7          \\
32    & Ge      & $\mu$g/g & 42    &         & 42       & 13         \\
33    & As      & $\mu$g/g & 15    &         & 15       & 4          \\
34    & Se      & $\mu$g/g & 32    &         & 32       & 8          \\
35    & Br      & $\mu$g/g & 81    &         & 81       & 1          \\
36    & Rb      & $\mu$g/g & 6     &         & 6        & 2          \\
37    & Sr      & $\mu$g/g & 16    &         & 16       & 10         \\
39    & Y       & $\mu$g/g & 2     &         & 2        & 2          \\
40    & Zr      & $\mu$g/g & 18    &         & 18       & 7          \\
$>$40 &         & $\mu$g/g &       &         &          & 16         \\ \midrule
\multicolumn{2}{l}{Total (\%)}          &      & 46.4  & 78.0    & 91.3     & 100.0      \\ \bottomrule
\end{tabular}
}
\caption*{Column 4 to 6 are derived from \cite{Art_Arndt_1996} and \cite{Art_Schramm_1989} with the sixth column representing the arithmetic mean of both studies. Column 7 is the meteorite composition, taking into account the meteorite portion of each group (Table \ref{TAB2}) and the respective meteorite group elemental mass abundances. For further details on the derivation see the text and \ref{APP2}.
\newline
\textsuperscript{a} The abundance of hydrogen is estimated as described in the text.}
\end{table}

By contrast, large meteoroids are mostly of asteroidal origin \citep{Art_Bottke_2002, InCol_Binzel_2015}, only a small portion originates from comets \citep{Art_Binzel_2004, Art_Fernandez_2005, Art_DeMeo_2008} or differentiated bodies (\citealp{NHM_catalogue}; \citealp[p. 258]{InCol_Borovicka_2015}; \citealp[p. 419]{InCol_Russell_2015}). A significant amount of material can survive upon entry and reaches the ground as meteorites. Therefore, we estimate the average elemental mass abundance of large meteoroids by calculating the composition of meteorites found on Earth. Meteorites are divided into different classes, based on their mineralogy and thus elemental composition. By weighting the composition of each meteorite class with their respective frequency of finds and falls on Earth, we calculate an average meteorite composition. Due to the large amount of classified meteorites of more than 22,000, this statistical approach is possible. We use the data given in \cite{NHM_catalogue} to yield the frequency of each meteorite class (see Table \ref{TAB2}). For the average elemental mass abundances we use data from \cite{Book_Meteorites_1974, Book_PlScCo_1998, inCol_Mittlefehldt_2001, Art_Demidova_2007}. The derived average composition of each meteorite class as well as details on the data are provided in \ref{APP2}. By weighting the elemental mass abundances with the class frequencies of Table \ref{TAB2}, we yield the overall elemental mass abundance of meteorites listed in Table \ref{TAB1}.

\begin{table}[!ht]
\centering
\caption{Frequencies of meteorite classes derived from finds and falls \citep{NHM_catalogue}.}
\label{TAB2}
\footnotesize{
\begin{tabular}{@{}llll@{}}
\toprule
\multicolumn{2}{l}{\textbf{}}    & \textbf{Class}                    & \textbf{Portion (\%)}          \\ \midrule
\multicolumn{3}{l}{\textbf{Chondrites}}                      & \textbf{91.95} \\
      & \multicolumn{2}{l}{\textbf{Ordinary Chondrites}}     & \textbf{86.35} \\
      &                      & H                             & 42.27          \\
      &                      & L                             & 37.72          \\
      &                      & LL                            & 6.36           \\
      & \multicolumn{2}{l}{\textbf{Carbonaceous Chondrites}} & \textbf{3.40}  \\
      &                      & CH                            & 0.08           \\
      &                      & CI                            & 0.04           \\
      &                      & CK                            & 0.54           \\
      &                      & CM                            & 1.18           \\
      &                      & CO                            & 0.62           \\
      &                      & CR                            & 0.57           \\
      &                      & CV                            & 0.36           \\
      & \multicolumn{2}{l}{\textbf{Enstatite Chondrites}}    & \textbf{1.22}  \\
      &                      & EH                            & 0.93           \\
      &                      & EL                            & 0.28           \\
      & \multicolumn{2}{l}{\textbf{Other Chondrites}}        & \textbf{0.99}  \\
      &                      & K (Kakangari)                 & 0.13           \\
      &                      & R (Rumurutiite)               & 0.85           \\ \midrule
\multicolumn{3}{l}{\textbf{Achondrites}}                     & \textbf{3.69}  \\
      &                      & Acapulcoites                  & 0.07           \\
      &                      & Angrites                      & 0.02           \\
      &                      & Aubrites                      & 0.28           \\
      &                      & Brachinites                   & 0.04           \\
      &                      & Lodranites                    & 0.09           \\
      &                      & Ureilites                     & 0.56           \\
      &                      & Winonaites                    & 0.07           \\
      & \multicolumn{2}{l}{\textbf{From Vesta\textsuperscript{a}}}              & \textbf{2.36}  \\
      &                      & Diogenites                    & 0.57           \\
      &                      & Eucrites                      & 1.22           \\
      &                      & Howardites                    & 0.57           \\
      & \multicolumn{2}{l}{\textbf{Lunar}}                   & \textbf{0.11}  \\      
      &                      & Lunaite                       & 0.11           \\
      & \multicolumn{2}{l}{\textbf{Martian}}                 & \textbf{0.09}  \\      
      &                      & Shergottites                  & 0.03           \\
      &                      & Nakhlites                     & 0.03           \\
      &                      & Chassignites                  & 0.03           \\
 \midrule
\multicolumn{3}{l}{\textbf{Stony Irons}}                     & \textbf{0.52}  \\
      &                      & Mesosiderites                 & 0.29           \\
      &                      & Pallasites                    & 0.22           \\ \midrule
\multicolumn{3}{l}{\textbf{Irons}}                           & \textbf{3.85}  \\
      &                      & IAB                           & 0.76           \\
      &                      & IC                            & 0.06           \\
      &                      & IIAB                          & 0.60           \\
      &                      & IIC                           & 0.05           \\
      &                      & IID                           & 0.09           \\
      &                      & IIE                           & 0.11           \\
      &                      & IIF                           & 0.03           \\
      &                      & IIIAB                         & 1.34           \\
      &                      & IIICD                         & 0.24           \\
      &                      & IIIE                          & 0.08           \\
      &                      & IIIF                          & 0.04           \\
      &                      & IVA                           & 0.37           \\
      &                      & IVB                           & 0.08           \\ \midrule
\multicolumn{3}{l}{\textbf{Total}}                           & \textbf{100}       \\ \bottomrule
\end{tabular}
}
\caption*{\textsuperscript{a} Expected to originate from Vesta.}
\end{table}

The metal (45\,\%) and metalloid (17\,\%) elemental abundance is higher than in IDPs, while the non-metallic portion (38\,\%) is lower. All in all, IDPs and meteorites show considerable differences, but are similar in the abundance of some elements.

The above described method of determination of the composition of the large meteoroids is biased, as meteoroids show different ablation rate and behaviour depending on their composition upon entry into the atmosphere. Thus, the amount of produced meteorites and the final composition to some extend depends on the properties of the initial meteoroids. Additionally, some meteorite classes are easier to find, e.\,g. iron meteorites are easier to distinguish from the environment due to their metallic look. Also, the number of meteorite finds and falls for some groups is quite low and the statistical sample might be insufficient in that case. Thus, there are uncertainties in our approach. 

\subsection{Atmospheric processing of the natural material} \label{sec:NATABL}
Results from various studies are used to derive ablation products of IDPs and larger meteoroids. Three ablation products are thought to be important: material due to deposition in the atmosphere in form of atoms, ions or molecules; material deposited as aerosols, e.g. particles of microns to nm size; material directly reaching the ground, thus not contributing to atmospheric injection.

\subsubsection{Small meteoroid ablation}
For small meteoroids, several studies suggest that there is a cutoff size below which no ablation is taking place, due to insufficient heating of the particle. This cutoff size ranges between meteoroid masses of 10$^{-11}$ to 10$^{-15}$\,kg depending on the study \citep{Art_Jones_1966, Art_Nicol_1985, Art_Popova_2004, Art_Vondrak_2008}. Meteoroids below this size can be treated as part of the aerosol mass fraction as they are slowed down to cm/s velocities. It takes them weeks to years to reach the ground depending on their size \citep{Art_Kasten_1968, Art_Rietmeijer_1998_Leonid, inCol_Rietmeijer_2002}. 

\begin{figure}[!th]
\begin{center}
\includegraphics[width= 1\columnwidth]{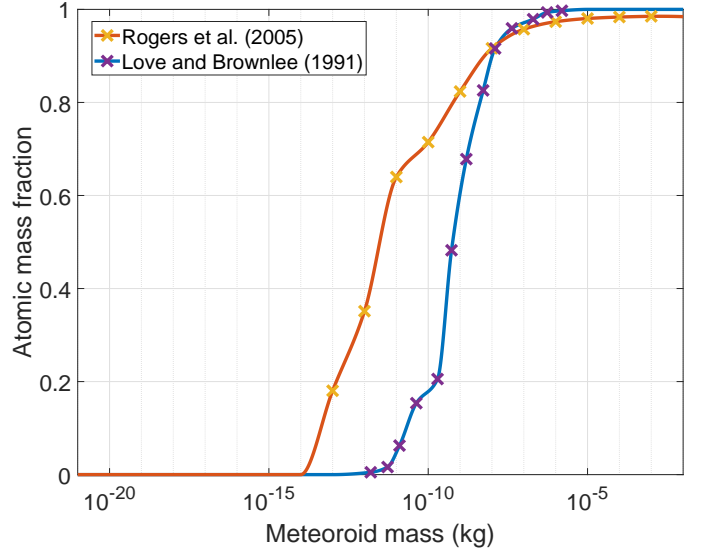}
\end{center}
\caption{Fraction of mass ablated in atomic form in dependence of meteoroid mass. The crosses depict the data points derived from the studies of \cite{Art_Rogers_2005} and \cite{Art_Love_1991}, while the line is the interpolation.}
\label{fig2}
\end{figure}

Looking at masses higher than the cutoff mass, there are three models to be considered. \cite{Art_Rogers_2005} present a numerical model of the ablation of small meteoroids in the mass range $10^{-3}$ to 10$^{-13}$\,kg for discrete velocities and different meteoroid densities. With the knowledge of the velocity distribution of meteoroids in the respective size range, an overall estimate of the ablation can be made. \cite{Art_Taylor_1996} state that the velocity distribution of meteoroids from $10^{-2}$ to $10^{-15}$\,kg is similar. We use the velocity distribution from \citet{Art_Taylor_1995}, tabulated in \cite{Tech_ECSS_2008}, and recalculate the distribution to an incident height of 100\,km, thereby taking into account the acceleration due to Earth's gravity. Values from \cite{Art_Rogers_2005} are weighted with that velocity distribution. To come closest to an IDP density of 2,200\,kg/m$^3$ \citep{Art_Carrillo_2016}, the mean of the ablated mass for two different particle densities of 1,000 and 3,300\,kg/m$^3$ is taken. Thus, we yield a final value for the fraction of ablated mass for the different mass bins (see Figure \ref{fig2}). Small meteoroids with masses above 10$^{-8}$\,kg show more than 90\,\% ablated mass. Towards lower masses, the fraction of ablated material decreases to almost zero. For this model, we adopt a cutoff size of 10$^{-14}$\,kg and interpolate the data (also shown in Figure \ref{fig2}). By weighting with the \cite{Art_Gruen_1985} mass influx, an ablated mass fraction of 86\,\% is obtained. We assume that all of the ablated material enters the atmosphere in atomic form, as recondensation of the vaporized material to dust is unlikely due to the small particle masses.

\cite{Art_Love_1991} have performed a similar study, simulating the atmospheric entry of over 50,000 meteoroids. Using their data on the amount of vaporized mass, equal to the mass of atoms ablated, and transforming the particle diameter to mass by using a density of 3,000\,kg/m$^3$, we yield the values and interpolation depicted in Figure \ref{fig2}. Here, we adopt a cutoff mass of 10$^{-13}$\,kg. Weighting with the \cite{Art_Gruen_1985} mass influx, an ablated mass fraction of 69\,\% is obtained, which is  considerably lower than the \cite{Art_Rogers_2005} estimate. Both, \cite{Art_Love_1991} and \cite{Art_Rogers_2005} show a very small to zero survivability of particles in the mass range 10$^{-6}$ to 10$^{-2}$\,kg. This is supported by \cite{inCol_Rietmeijer_2002}.

A third study, \cite{Art_Carrillo_2016}, utilizes the chemical ablation model CABMOD, introduced by \cite{Art_Vondrak_2008} incorporating differential ablation of different elements along with the model of the zodiacal cloud \citep{Art_Nesvorny_2010, Art_Nesvorny_2011}. They use a different mass distribution and a different velocity distribution with a lower mean velocity than used in our study. For their mass range of 10$^{-6}$\,kg to 10$^{-12}$\,kg, only 18.2\,\% of the material are ablated atoms. Taking the same mass range, the two other study interpolations yield a fraction of 85\,\% and 65\,\%, respectively. Thus, differences to the other two models are large, partly to be explained by the slower average velocity and the different mass distribution. In the following we use the simulated values of \cite{Art_Love_1991}. 

\subsubsection{Large meteoroid ablation}

\begin{figure*}[!th]
\begin{center}
\includegraphics[width= 1.8\columnwidth]{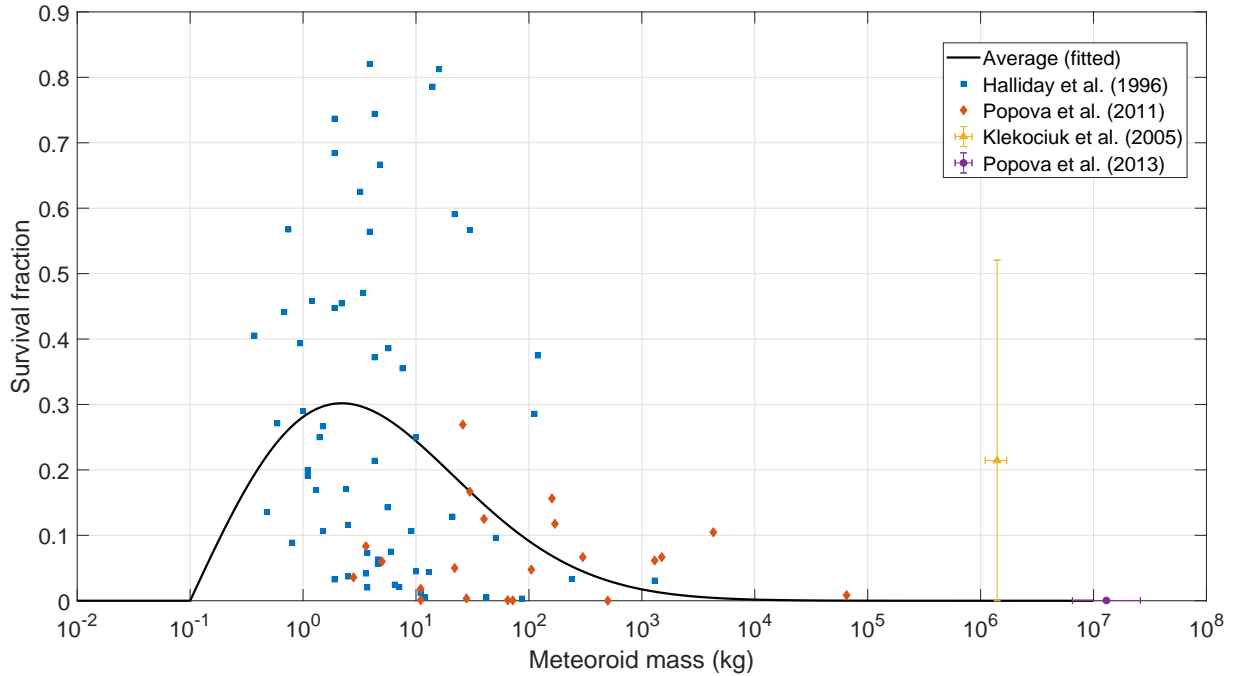}
\end{center}
\caption{Survival rate of large meteoroids. Shown are fireball data from different studies. Errors are depicted if available. The black line depicts our fitted estimate; for further details see the text.}
\label{fig3}
\end{figure*}

For large meteoroids, ablation largely reduces the mass of entering meteoroids. However, a substantial fraction can survive atmospheric entry (\citealp[p. 236]{inCol_Rietmeijer_2002}; \citealp[p. 258]{InCol_Borovicka_2015}). The survival fraction, the ratio of the meteoroid's terminal and initial mass, is highly dependent on velocity, density, composition and initial mass itself. The dependence of the survival fraction on mass is displayed in Figure \ref{fig3}, which is based on results of \cite{Art_Halliday_1996, Art_Klekociuk_2005, Art_Popova_2011, Art_Popova_2013}. Data are widely scattered due to the various dependencies mentioned above. An average mass dependent survival fraction is derived by fitting a scaled Rayleigh distribution to all the given data. The starting point of the distribution is chosen to be at 10$^{-1}$\,kg with no mass survival from 10$^{-2}$ to 10$^{-1}$\,kg considering \cite{Art_Baldwin_1971}. This fits the results by \cite{Art_Love_1991, inCol_Rietmeijer_2002, Art_Rogers_2005} and also matches our model of the small meteoroid ablation. The resulting survival fraction model $S(m)$ with 

\begin{equation}
    S(m) = \left\{%
\begin{array}{ll}
    0 &, \hbox{$-2\leq\log{m}<-1$} \\
    0.67\,\frac{\log(m)+1}{1.34^2}e^{-\left( \frac{\log(m)+1}{2\cdot 1.34}\right) ^2}& , \hbox{$-1\leq \log{m}\leq8$} \\
\end{array}%
\right.
\end{equation}
where $S$ denotes the survival fraction, and $\log{m}$ the decadic logarithm of the meteoroid initial mass, is displayed in Figure \ref{fig3}.

In a further step one needs to clarify how much of the ablated material is deposited in the atmosphere in atomic or aerosol form. Here, ablated material can recondense to dust and also dust particles can leave the fireball as it is ablated. Observations of dust clouds are very rare and incorporate large errors. The Chelyabinsk object created a dust cloud of roughly 24\,\% of the initial mass of approx. $1.3\cdot 10^{7}$\,kg \citep{Art_Popova_2013}. \cite{Art_Klekociuk_2005} provide observational results from an entry of a massive meteoroid (roughly $1.4\cdot 10^{6}$\,kg). The dust cloud amounted to roughly 79\,\% of the total meteoroid mass, at least 47\,\%. TC$_3$, a smaller bolide of around $5\cdot 10^{4}$\,kg produced a dust cloud of about 20\,\% (at least 15\,\%) of the initial meteoroid mass \citep{Art_Borovicka_2009}. Detailed modelling of a fireball entry by \cite{Art_Borovicka_2019} indicates that more fragmentation leads to more dust being released. This would point towards an increase in the dust fraction for larger meteoroids as fragmentation events are more likely for larger meteoroids. Therefore, we assume a linear increase in the aerosol fraction with increasing logarithm of mass. For meteoroid masses of 10$^{-2}$\,kg, all the material is ablated in atomic form in accordance with the findings in the previous section. The aerosol mass fraction increases to 50\,\% for a mass of $10^{7}$\,kg.

\subsubsection{Resulting mass dependent ablation}
\begin{figure*}[!th]
\begin{center}
\includegraphics[width= 1.8\columnwidth]{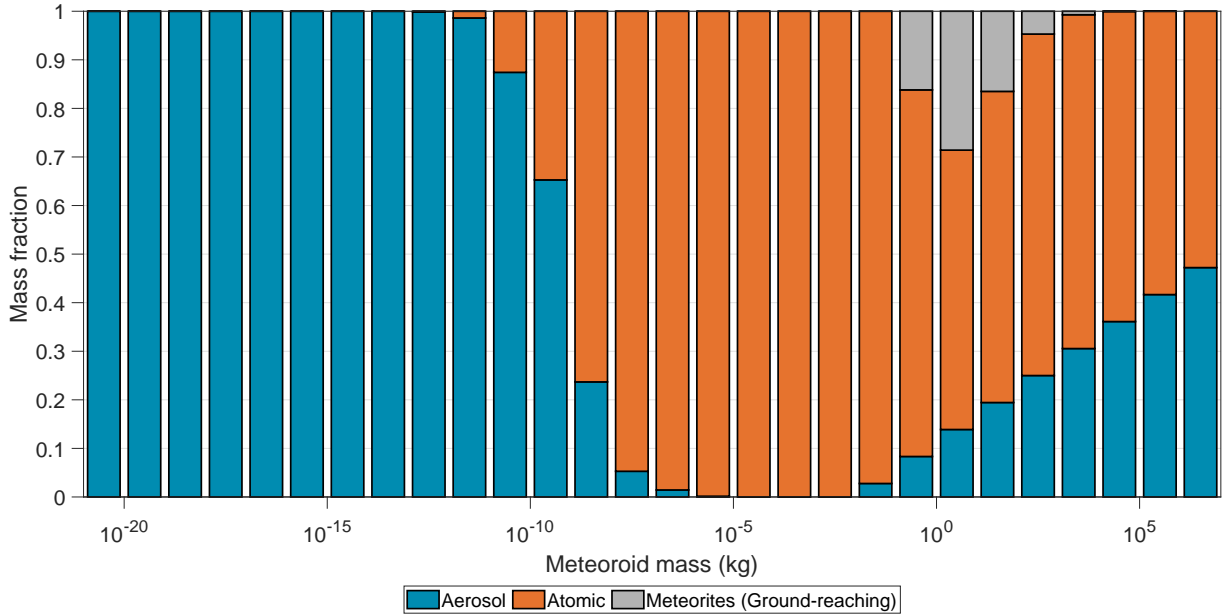}
\end{center}
\caption{Estimated fractional mass ablation of meteoroids in dependence of the meteoroid mass. Values are given for each mass decade. For details on the derivation, see the text. Note that all of those values are rough estimates.}
\label{fig4}
\end{figure*}

The mass dependent fraction of the three different ablation products is displayed in Figure \ref{fig4}, based on the results from the previous two sections. For nearly every meteoroid mass, the material is injected into the atmosphere either in atomic or aerosol form. Only for a very limited mass range significant amounts of the entering meteoroids reach the ground directly upon entry into the atmosphere.

\subsection{Overall natural injection} \label{sec:NATINJ}
With the estimates of the mass distribution, the composition, and the ablation of incoming meteoroids a complete picture of the injection of natural matter into Earth's atmosphere is available. The following estimates emerge. 12,325\,t natural material are entering Earth's atmosphere every year. Only 48\,t/yr of meteoroids are reaching the ground (0.4\,\% of the whole mass) directly upon entry. The rest is injected into the atmosphere, 8,421\,t (68\,\% of the whole mass) in atomic form, 3,904\,t (32\,\%) as aerosols. Most of the material is non-metallic (5,674\,t), but metals are also significant (4,047\,t). Metalloids take the smallest portion (1,655\,t). The most abundant metals are iron (2,295\,t) and magnesium (1,300\,t), other metals only contribute with minor fractions, e.\,g. aluminum (131\,t), nickel (90\,t), calcium (88\,t) and sodium (59\,t). Non-metallic and metalloid elements with high injection masses are oxygen (3,851\,t), silicon (1,654\,t), carbon (1,054\,t), sulfur (513\,t) and hydrogen (220\,t).


\section{Anthropogenic injection}\label{sec:ANTMAT}
Since the beginning of the space age, anthropogenic injection into the upper atmosphere occurs. Decommissioned spacecraft, rocket bodies, and other debris are entering Earth's atmosphere. This is due to the aerodynamic drag of the atmosphere, which is reducing the speed of orbiting objects even at altitudes as high as 1000\,km. Most of the space components are made of metals. 

With the ongoing use of space, more and more space debris is inserted into orbits around Earth. Space debris has become a serious problem as it is a hazard to operating spacecraft and even the International Space Station. All the debris in LEO can lead to a cascade effect of debris impacting satellites causing even more debris and so forth, possibly rendering whole orbits unusable for decades to hundreds of years. In order to reduce the amount of future space debris and ensure safety in space, guidelines have been introduced \citep{Tec_ESA_SDM, Tec_ISO24113:2019, Tec_NASA-STD-8719.14}. For LEO, satellites and upper stages have to be de-orbited within 25 years after their end of lifetime. Due to the increasing use of space, these requirements might tighten in the following years and decades. Thus, nearly everything launched into LEO nowadays will burn up in the atmosphere, eventually.

\subsection{Today's annual anthropogenic mass influx}\label{sec:ANTMASS}
We derive an estimate of the annual mass influx from the altitude depending mass distribution around Earth as provided by \citet{Art_ODQN_Liou_2018}. Due to the atmospheric drag, there is an altitude below which, on average, all objects reenter the atmosphere within one year. With that, the mass entering every year is the sum of the mass of all objects below this altitude. Using \cite{Art_Boykin_1966} together with data from \cite{Art_Bowman_2002,Art_Saunders_2012}, we come up with an average reentry altitude of 450\,km, roughly matching calculations by \cite{InPro_Braun_2013}. As a result, the annual mass influx amounts to about 190\,t/yr, a value comparable to that one estimated by \citet{InPro_Pardini_2013}. About 60\,\% of the mass are spacecraft, 40\,\% rocket bodies. This mass influx value will increase in the future as the amount of mass in orbit is rising continuously \citep{Art_ODQN_Liou_2020}. It should be noted that space debris is large in numbers but contributes only a negligible part to the anthropogenic mass influx. 

In a further step, the core stages of launch vehicles need to be considered. Although these are accelerated to considerable velocities, they remain suborbital and reenter the atmosphere right after liftoff. These objects are not tracked by space agencies and therefore are not included in the studies by \cite{InPro_Pardini_2013} and \cite{Art_ODQN_Liou_2018}. To estimate the mass contribution by core stages, we consider the launch history of 2019. From each orbital launch, we consider every rocket stage that is jettisoned into a suborbital trajectory. Using available data on launch profiles, the approximate entry velocity of each stage can be calculated. Only stages with entry velocities higher than 3.8\,km/s are taken into account as we consider lower entry velocities to be insufficient for significant ablation and contribution to the injection rate into the atmosphere. Additionally, we neglect launch vehicles with a payload mass lower than 1\,t as well as suborbital rocket launches. A complete list of the data and sources is given in \ref{APP3}. Using the mass of each stage, we estimate about 702\,t of rocket stage mass reentering Earth's atmosphere in 2019, with speeds from 3.8 to \,km/s to 7.6\,km/s. This mass value will increase in the near future, too. Summarizing, today's (2019) annual mass influx from anthropogenic sources amounts to about 890\,t. The largest contribution (87\,\%) are rocket bodies.

\subsection{Satellite constellations} \label{sec:SATCONST}
\begin{table*}[!ht]
\caption{Satellite constellation projects with more than 100 satellites and satellite masses greater than 10\,kg.}
\label{TAB3}
\centering
\small{
\renewcommand{\arraystretch}{1.6}
\begin{tabular}{%
>{\raggedright\arraybackslash}p{2cm}%
>{\raggedright\arraybackslash}p{1.5cm}%
>{\raggedright\arraybackslash}p{1.4cm}%
>{\raggedright\arraybackslash}p{1.8cm}%
>{\raggedright\arraybackslash}p{1.4cm}%
>{\raggedright\arraybackslash}p{4.2cm}%
>{\raggedright\arraybackslash}p{3.1cm}}
\toprule
Project  & Country & Satellite mass (kg)              & Proposed sat. number        & Height (km) & Project Status                & Sources 
\\ 
\midrule

Starlink (SpaceX)       & USA     & 260                               &  \textbf{7518}    \newline   \textbf{1548}+2824\textsuperscript{a} \newline \textit{30000} & 335--346 \newline 540--570 \newline  300--600 & 538 launched, of that 60 prototypes \citep{Misc_starlink09, Misc_starlink1}, filing for 30000 more satellites to FCC prepared \citep{Art_Koksal_2019}  & \cite{Tech_FCC_18-161, Tech_FCC_18-38, Tech_DA_19-342, Tech_SpaceX_SAT-MOD-2020}       
\\
OneWeb           & UK  & 147                               & \textbf{720}\textsuperscript{b} \newline 1280 \newline \textit{1280}\textsuperscript{c} \newline \textit{47128}        & 1200 \newline 8500 \newline 8500 \newline 1200         & 74 satellites launched  \citep{Tech_Arianespace_OneWeb}, LEO extension to 47844 satellites requested in 2020   &  \cite{Tech_FCC_17-77, Misc_FCC_Withdraw_SAT-AMD-2018, Tech_OneWeb_SAT-AMD-2018, Tech_OneWeb_SAT-LOI-2017, Tech_OneWeb_SAT-MOD-2020}      
\\
Telesat LEO (Telesat)    & Canada  & 168\textsuperscript{d}  & \textbf{117}+292 & 1000       & 1 prototype launched        & \cite{inPro_Grant_2019, Misc_FCC_PressRelease_2018}       
\\
Kepler (Kepler Communications) & Canada     & 12--15                           & \textbf{140}                         & 500--600              & 2 prototypes (3U Cubesats) launched  & \cite{Misc_FCC_PressRelease_2018, inPro_Grant_2019}      
\\
Project Kuiper (Amazon) & USA     &                                 & 3236                         & 590--630              & FCC filing submitted in 2019  & \cite{Tech_Kuiper_SAT-LOA-2019}     
\\
Hongyun (CASC)   & China   & 247\textsuperscript{d} & 320                     & 1000       & 1 prototype launched          &  \cite{inPro_Grant_2019}       
\\
Xingyun (CASIC)   & China   & 93 & 156                      & 570   & 2 satellites launched in 2020          &  \cite{inPro_Grant_2019}       
\\
Satellogic     & Argentina         &  37  & 300       &     500          &  8 satellites launched since 2016 & \cite{Book_Lal_2017} 
\\
Boeing       & USA     &                                  & \textit{3000}        & 1200             & Boeing withdrew several FCC filings and   & \cite{inPro_Grant_2019, Book_Lal_2017}     
\\ 
Samsung  & South Korea  &                                  & \textit{4600}                         & 300--2000       & No perceptible action since 2015 & \cite{Misc_Khan_2015}        
\\
MCSat (Thales) & France     &                          & \textit{800}+\textit{4000}                  &               & Beside filings at the International Telecommunication Union (ITU), no information available  & \cite{inPro_Grant_2019} 
\\
3ECOM-1 & Lichtenstein     &                          & \textit{288}                  &               & Beside ITU filings, no information available  & \cite{inPro_Grant_2019} 
\\
\bottomrule
\end{tabular}
\renewcommand{\arraystretch}{1}
}
\caption*{The list might not be complete as companies in this area emerge and disappear rapidly. Reliable information are hard to gather. The font used for the proposed satellite number (column 4) indicates the probability of realization: an italic font means that the realization is not safe or improbable, normal font implies a high probability for the project realization, and a bold font means that the constellation is granted by the FCC of the USA and will therefore most likely be materialized or is already in process. Additional information is acquired from company websites and satellite launch data. The projects are ordered after their probability of realization and the number of spacecraft.\newline
\textsuperscript{a} SpaceX currently has approval for 1,548 satellites at 550\,km altitude and 2,825 satellites at 1,110--1,325\,km altitude. They have filed for a modification of the orbit altitude to 540--570\,km and reduction to 2,824 satellites. As earlier modifications of this kind were successful, it is likely this gets granted.\newline 
\textsuperscript{b} OneWeb has authorization of the launch of 720 satellites at this altitude but company statements suggest only 648 are needed.\newline
\textsuperscript{c} OneWeb filed at the FCC for a doubling of that number, but has withdrawn that request.\newline 
\textsuperscript{d} Prototype mass.}
\end{table*}

With evermore companies engaging in commercial spaceflight, satellite constellation projects have been proposed and some of them already started. Mostly, these projects aim at providing global telecommunication services for the global internet \citep{Art_ODQN_Liou_2018}. Therefore, hundreds to thousands of satellites used as relays will be brought to LEO and eventually, after reaching their end of lifetime, burn up in the atmosphere. In Table \ref{TAB3}, proposed and (partially) realized large satellite constellation projects are listed with additional information such as characteristics of the satellites or the current status. In total, constellations of nearly 110,000 satellites have been proposed. The two constellations of SpaceX and OneWeb are about to operate soon \citep{Tech_Arianespace_OneWeb, Misc_starlink1}, others will most certainly follow.

Constellation satellites are launched into relatively low orbits because of the effective range of antennas and latency. Some satellites even have to raise and retain their orbit by themselves using on-board propulsion. Most likely, due to the large number of satellites, some of the satellites will fail in orbit due to electronic failure or propulsion problems. From the first Starlink launch, 3 out of 60 satellites did not seem to work properly. Thus, we assume a failure rate of 5\,\% for the satellites, which have to be replaced. This increases the mass estimate. Additionally, we expect most of the satellites at high orbits (around 1,000\,km) to be de-orbited after service, as many companies have already vowed to do so. 

In order to estimate the total mass influx caused by constellations, upper stages and core stages of the launching rockets have to be incorporated, too. Depending on the launch vehicle, the upper stage mass relative to the payload mass is different. Typical launch vehicles are the Falcon 9 and the Soyuz 2.1 Fregat rockets. The Falcon 9 payload is 15.6\,t \citep[derived from][]{Misc_starlink1}, that one of the Soyuz 2.1 Fregat 5\,t \citep{Tech_Arianespace_OneWeb}. With their respective upper stage and core stage mass \citep{Datasheet_NASA_Falcon9, Tech_RS_Soyuz} one can estimate a typical ratio of payload to upper (core) stage mass of 0.24\,t (0.89\,t) per ton launched and reentering within years. This is included into the subsequent calculations. Based on statements from some companies, we expect the lifetime for each satellite to be around 5 years. This means, every 5 years a whole constellation has to be replaced. 

In the following, we consider two scenarios for a possible future mass influx:

\subsubsection{Future Scenario 1 --- Probable influx}
This scenario assumes today's, 2019 based mass influx. To this influx we add that one due to satellite constellations most probably installed in the future (constellation projects from Table \ref{TAB3} with bold and normal font type number of satellites). We expect all satellites in LEO to reenter Earth's atmosphere. However, we do not expect the satellites of OneWeb's 8,500\,km height constellation (see Table \ref{TAB3}) to enter the atmosphere; only their upper and core stages will reenter and are taken into account. All in all, in 5 years additional 19,411 satellites as well as 585 upper stages and 443 core stages will be brought to orbit. With the above mentioned lifetime, failure rate and upper stage and core stage per payload mass, every year, 960\,t of  satellites, 291\,t upper stages, and 1,491\,t core stages will reenter the atmosphere. All in all, the annual mass influx amounts to 2,742\,t/yr. However, a significant portion of that material may reach the ground, as discussed further down. 

\subsubsection{Future Scenario 2 --- Maximum influx}
This scenario assumes a doubling of the 2019 mass influx. To this influx all of the planned satellite constellation projects are added. However, from the projected additional 30,000 Starlink and nearly 48,000 OneWeb satellites we merely assume 50\,\% of them being realized. Thus, within 5 years, nearly 75,000 additional satellites, 1,984 upper stages, and 1,503 core stages are considered in this scenario. The total mass influx is estimated at 8,114\,t/yr, consisting of 3,153\,t satellites, 880\,t upper stages, and 4,081\,t core stages. 

\subsection{Composition of the anthropogenic material} 
The composition of satellites and rocket bodies differs largely from the composition of meteoroids. Generally, the metal abundance is way higher as most structural components are made of alloys. We distinguish between rocket bodies (core stages that reenter right after launch and upper stages which reach orbit) and spacecraft. 

Rocket bodies mainly consist of propulsion tanks and rocket engines. For composition information of the propulsion tanks we use the information provided by \citet{InCol_Henson_2018}. Tanks have to withstand the pressure of the loaded fuel and thus are made out of durable alloys. Most common is AA2219 aluminum alloy but newer rockets like SpaceX' Falcon 9 are also made out of Al-Li alloys like AA2198 \citep{InCol_Wanhill_2014}. On the other hand, the Centaur tank is manufactured out of 301 stainless steel. Those rockets using solid rocket motors (e.\,g. Antares or Vega launch vehicles) use D6AC steel or similar for their tanks. Other parts like feedlines, other pressured vessels, and unpressurized structure are made out of steels, Al-alloys and Ti-alloys. Non-metallic materials are rare. They are, for example, found in thermal insulation and adapters. Based on the available information and respecting the large variety of materials we assume the following relative composition values as a first estimate:  80\,\% AA2219 Al-alloy, 5\,\% AA2198 Al-Li-alloy, 5\,\% D6AC steel, 5\,\% 301 steel, 5\,\% others. 

Liquid fuelled engines make up the other part of a rocket body. They need even more resistant alloys, especially due to the high temperature load in combustion chambers. According to \citet{InCol_Halchak_2018}, different groups of alloys are used. For the nozzle, the most heavy part of an engine, Ni-alloys like Inconel 718 (used in the Space Shuttle main engine RS-25) or Inconel 600 (Vulcain 2) are the most common alloys. The much used RD-107 and RD-108 engines of the Soyuz launch vehicle incorporate Copper-Chromium alloys (3\,\% Cr). A similar alloy called Narloy-Z has been used in the combustion chamber of the RS-25. RD-107 and RD-108 nozzles are additionally made of stainless steel, nickel-based alloys or other metals. Other components of engines like propellant pumps, components of turbopumps, valves, and feed lines are made of different alloys, typically Al-, Ti-, or Ni-based due to the required durability. Only a minor portion is made of non-metallic components like silicon carbide. Again, based on the variety of different materials being in use, we use the following estimate for the composition of the rockets' engines: Inconel 718 Ni-alloy, Inconel 600 Ni-alloy, and A286 alloy each 25\,\%, 10\,\% Cu-Cr-alloy (alloy  with 3\,\% Cr), and 5\,\% Ti, 5\,\% Al, and 5\,\% Ni. 

We additionally differentiate between core stages and upper stages. For upper stages, the engine mass portion of the whole rocket body normally is lower than for core stage. Looking at the Ariane 5, Soyuz 2.1 Fregat, and Atlas V launch vehicles, we estimate 8\,\% mass (dry mass) of upper stages taken up by the engine, while for core stages it is 18\,\%. 

Satellites differ largely in composition and size depending on their field of use, mission lifetime, etc. \cite{InCol_Finckenor_2018} provides some insight into materials used in spacecraft. The structure mainly consists of Al-alloys, Ti-alloys, or stainless steel. Even Ni-alloys are in use, if more durability is needed. Aluminum is a preferred material as it is lightweight. Generally, most structures are made out of metals. However, some parts of the spacecraft, e.\,g. the outer hull exposed to the sun, are constructed out of material capable of withstanding thermal expansion. These materials are mostly non-metallic, for example polyimide, graphite or fiberglass. Non-metallic materials can also be found in orbital debris and meteoroid shielding, although Al-sheets are common. As many spacecraft in LEO are Earth observation satellites, optical materials like quartz or mirror material also play a role in the overall mass composition. A very important mass contributor of satellites are solar arrays. We estimate their mass contribution to be 16\,\%. We assume half of the solar arrays to be old-fashion Si solar arrays, whereas the other half use newer technology with multijunction solar cells made of layers of Ge, Ga-As, and Ga-In-P. All in all, we assume the following mass composition: 40\,\% Al, 5\,\% Ni, 5\,\% Ti, 10\,\% Fe, 8\,\% Si, 4\,\% Ge, and Ga, In, P, and As each 1\,\%. The remaining 24\,\% are other materials. We expect the large constellation satellites to also match this composition reasonably well. 

\subsection{Atmospheric processing of the anthropogenic material}
Spacecraft and upper stages entering the atmosphere have a much longer interaction time than meteoroids due to their shallower entry angle and small entry velocity of about 7\,km/s. This is due to these anthropogenic objects approaching almost circular orbits during the reentry phase. Therefore, their ablation is largely different from that one of meteoroids. Anthropogenic material reaches temperatures of 850 to 1950\,K \citep{inPro_Rochelle_1997, inPro_Ailor_2005,Art_Lips_2017} while meteoroids and fireballs can reach temperatures above 3000\,K \citep{Art_Borovicka_1993, Art_Jenniskens_2004}. This strong difference in temperature and the different materials imply different ratios of the atomic and aerosol ablation for meteoroids/fireballs and anthropogenic material.  In need of better data, we assume a higher aerosol fraction (75\,\%) for the anthropogenic material while that one of meteoroids is only about 30\,\% (compare with Figure \ref{fig4}). This is due to higher temperatures causing transition into the gas and/or plasma phase. 

Due to the lower ablation temperature and the high mass (in the order of tons) of the anthropogenic material, a significant fraction of their mass reaches the ground. Today, survival rates of human-made objects are expected to range between 5 to 40\,\% \citep{inPro_Ailor_2005, Art_Anselmo_2005, Art_Pardini_2019}. Simulations with reentry software indicate values in this range, too \citep{Art_Anselmo_2005, InCol_Klinkrad_2006, inPro_Kelley_2010}. The thermal ablation of spacecraft and upper stages differs as well due to the different structure. For spacecraft, we assume an average survivability of 20\,\%, for upper stages 35\,\%, and for core stages 70\,\%. Large constellation satellites are estimated to burn up completely in the atmosphere \citep{Tech_SpaceX_SAT-LOA-2016, Tech_SpaceX_SAT-MOD-2018, Tech_OneWeb_SAT-LOI-2016}.

\subsection{Overall anthropogenic injection} \label{sec:ANTINJ}
Combining all the information about the annual anthropogenic mass influx, composition, and atmospheric processing provides the following estimates for today's injection and the two different future scenarios emerge. 

\begin{table*}[!bht]
\centering
\caption{Anthropogenic and natural injection for the different ablation products. Masses are given in t/yr. Numbers in parenthesis are the percentage compared to the value of the natural material in the respective column.}
\label{TAB4}
\small{
\begin{tabular}{@{}ll|p{0.5cm}lp{0.5cm}lp{0.5cm}l|p{0.5cm}l@{}}
\toprule
               &            & \multicolumn{2}{l}{Atomic} & \multicolumn{2}{l}{Aerosol} & \multicolumn{2}{l}{Total injection} & \multicolumn{2}{|l}{Ground-reaching} \\ \midrule
\multirow{3}{*}{Anthropogenic}        & Today    & 88 & (1.0)
  & 263  & (6.7)  & 351 & (2.8)   &  541 & (1,138)       \\
  & Scenario 1 & 393 & (4.7)  & 1,180 & (30.2)   & 1,573 & (12.8)  & 1,168 & (2,458)       \\
  & Scenario 2 & 1,226 & (14.6)  & 3,678 & (94.2)   & 4,904 & (39.8)  & 3,210 & (6,753)       \\ \midrule
Natural  &           & 8,421 &  & 3,904 &   & 12,325 & & 48 &         \\ \bottomrule
\end{tabular}
}
\end{table*}

\begin{table*}[!bht]
\centering
\caption{Anthropogenic and natural injection per element group. Masses are given in t/yr. The numbers in parenthesis depict the percentage compared to the natural injection value in the respective column.}
\label{TAB5}
\small{
\begin{tabular}{@{}cl|p{0.5cm}lp{0.2cm}lp{0.1cm}llp{0.5cm}l@{}}
\toprule
                &            & \multicolumn{2}{l}{Metals} & \multicolumn{2}{l}{Metalloids} & \multicolumn{2}{l}{Non-metals} & Not assignable & \multicolumn{2}{l}{Total injection} \\ \midrule
\multirow{3}{*}{Anthropogenic}        & Today    & 305 & (7.5)  & 12 & (0.7)      & 1 & (0.02)     & 33  & 351 & (2.8)      \\
                                      & Scenario 1 & 1,189 & (29.4)  & 123 & (7.4)      & 10 & (0.2)      & 252 & 1,573 & (12.8)      \\
                                      & Scenario 2 & 3,643 & (90.0)  & 406 & (24.6)      & 32 & (0.6)      & 822  & 4,904 & (39.8)      \\ \midrule
\multicolumn{1}{l}{Natural injection} &            & 4,047 &   & 1,655 &       & 5,674 &       & 949  & 12,325 &      \\ \bottomrule
\end{tabular}
}
\end{table*}

\subsubsection{Today's influx}
Currently, 892\,t of anthropogenic material enters Earth's atmosphere every year of which 88\,t are injected in atomic form; aerosols make up 263\,t. The remaining material (541\,t) reaches the ground. From the injected elements, aluminum is most abundant with 211\,t, followed by iron (36\,t), nickel (23\,t), and copper (15\,t). Metals make up at least 86\,\% of the injected material. 

\subsubsection{Future Scenario 1}
For Scenario 1, the annual anthropogenic mass influx increases drastically to 2,742\,t. 1,573\,t are injected into the atmosphere, 1,180\,t as aerosols, 393\,t in atomic form. Again, aluminum is the largest part of the injection with 807\,t, followed by iron (159\,t), nickel (89\,t), and silicon (76\,t). Again, most of the injected material is metal (at least 75\,\%). 

\subsubsection{Future Scenario 2}
For Scenario 2, the annual anthropogenic mass influx increases even more to 8,114\,t from which 4,904\,t are injected into the atmosphere. Aerosols contribute 3,678\,t, material of atomic form 1,226\,t. The order of the most injected elements is the same as in Scenario 1 with aluminum (2,467\,t), iron (496\,t), nickel (272\,t), and silicon (251\,t). The metal portion is at least 74\,\%.

\section{Final results and comparison} \label{sec:COMPAR}

With all the available information, the natural and anthropogenic injection can be tabulated and compared. Three aspects have been evaluated in this study: the injection by ablation products (Table \ref{TAB4}), by element group (Table \ref{TAB5}), and the injection of selected elements (Table  \ref{TAB6}). Today, the injection into the atmosphere is dominated by natural material. About 2.8\,\% of the mass is of human origin. Although metals are highly abundant in spacecraft and rocket bodies, the anthropogenic metal injection is also well below the natural metal injection. However, there are elements which are injected mainly by human-made objects, for example aluminum or copper. The anthropogenic injection can also prevail the natural injection for some specific elements that are not very abundant in the solar system and therefore in meteoroids, e.\,g. germanium. 

With the incorporation of large satellite constellations, the injection situation changes strongly. The near future Scenario 1 predicts 1,573\,t of anthropogenic material injected into the atmosphere, which is already 12.8\,\% of the natural injection. For the extreme Scenario 2 we infer an anthropogenic mass injection rate of about 39.8\,\% of the natural rate. For metals, the injection is even higher with 29.4\,\% (Scenario 1) and even 90.0\,\% (Scenario 2) of the natural metal injection, respectively. Additionally, there are more elements for which the anthropogenic injection surpasses the natural injection: for example titanium  (2459\,\%), chromium (131\,\%), and nickel (304\,\%) for Scenario 2. Satellite constellations also lead to a massive enhancement of the injection of aluminum and copper. 

The anthropogenic injection also increases the injection of aerosols disproportionally as we estimate the entry of human-made objects to produce more aerosols than atoms. Today, human-made bodies make up 6.7\,\% compared to the natural injection, while atomic material is only at 1.0\,\%. For future satellite constellations, the aerosol fraction increases to 30.2\,\% and 94.2\,\% for the two scenarios.

\begin{table}[!ht]
\centering
\caption{Anthropogenic and natural injection of some selected elements. Masses are given in t/yr. The numbers in parenthesis depict the percentage compared to the natural injection value in the respective row. Note that percentages larger than 100\,\% indicate that these elements are mainly of anthropogenic origin. For some elements, no anthropogenic abundances were calculated. }
\label{TAB6}
\small{
\begin{tabular}{@{}l|p{0.3cm}lp{0.2cm}lp{0.4cm}l|r@{}l}
\toprule
 \multirow{2}{*}{El.}   & \multicolumn{6}{c|}{Anthropogenic} & \multicolumn{2}{l}{\multirow{2}{1.4cm}{Natural \newline injection}} \\
 &  \multicolumn{2}{l}{Today}     & \multicolumn{2}{l}{Scenario 1} & \multicolumn{2}{l|}{Scenario 2}    &       \\ \midrule
H   &   &        &  &          &          &     & 220&   \\
C   & 0.1 & (0)   & 0.2 & (0)    & 0.5 & (0)       & 1,054& \\
O   &  &         &       &     &    &           & 3,851& \\
Mg  & 0.04 & (0)  & 0.1 & (0)    & 0.3 & (0)       & 1,300& \\
Al  & 211 & (161) & 807 & (614)  & 2,467 & (1,877) & 131&   \\
Si  & 8 & (0)     & 76 & (5)     & 251 & (15)      & 1,654& \\
S   &    &       &      &      &        &       & 513&   \\
Ti  & 7 & (100)   & 52 & (754)   & 171 & (2,459)   & 7&     \\
Cr  & 7 & (20)    & 17 & (47)    & 48 & (131)      & 37&    \\
Fe  & 36 & (2)    & 160 & (7)    & 496 & (22)      & 2,295& \\
Ni  & 23 & (25)   & 89 & (99)    & 272 & (304)     & 90&    \\
Cu  & 15 & (720)  & 38 & (1,747) & 106 & (4,923)   & 2&     \\
Ge  & 4 & (776)   & 37 & (7,973) & 124 & (26,435)  & 0&.5   \\
\bottomrule
\end{tabular}
}
\end{table}

\section{Conclusion} \label{sec:CONC}
The extensive review, analysis, and estimates presented in this study provide an overview on the natural and anthropogenic injection of matter into Earth's atmosphere. At the present time, the anthropogenic injection already contributes a non-negligible amount of mass to the injection. With large satellite constellations, proposed and started from companies all over the world, the anthropogenic injection will become significant compared to the natural injection. Although many of the values used to estimate the injection inhibit uncertainties due to different scientific results on many topics or insufficient data, the results of this study should raise attention and also concern towards the alteration of Earth's atmosphere due to the reentry of human-made spacecraft and rocket bodies. Especially looking at metals, the anthropogenic injection may well exceed 30\,\% of the whole material deposited in the upper atmosphere every year. Overall, in the near future we need to be prepared that the injection of anthropogenic material will increase to 12.8\,\% -- 39.8\,\% of the natural injection. Those values clearly show that the anthropogenic injection is not negligible in the near future and requires further  consideration with respect to their impact on Earth's atmosphere.

The uncertainties involved demonstrate that more research needs to be done to clarify the significance of the effects of the human use of space on Earth's habitat. There are many different possible effects on the atmosphere that may be caused by an increased injection. For example, the large amount of aerosols injected by the ablation of anthropogenic material may have an effect on Earth's climate as aerosols in the high-altitude atmosphere have a negative radiative forcing effect \cite{Art_Lawrence_2018}. 

Beside the intensively discussed problem of space debris \citep[e.\,g.][]{Book_2006_Klinkrad} we conclude that the re-entry of human-made objects into the upper atmosphere may have a significant effect on our habitat and needs more attention in future studies. Advances in technology and a stronger and stronger use of Earth's environment always have side effects that are most often not perceived at the beginning of innovation and progress.

\section*{Acknowledgements}
The authors thank Carsten Wiedemann, Martin Sippel, Sven Stappert, Gerhard Drolshagen, and J{\"u}rgen Blum for helpful discussions.

\appendix
\section{}\label{APP1}
The annual mass influx per mass decade shown in Figure \ref{fig1} is derived separately for the three different mass ranges. For masses between $10^{-21}$ to $10^{-2}$\,kg, the interplanetary flux model from \cite{Art_Gruen_1985} with the flux at 1\,AU given as 
\begin{align}
    \nonumber F(m) & = && (2.2\cdot10^{3} \, m^{0.306}+15)^{-4.38} \\ 
    \nonumber & && + 1.3\cdot10^{-9} \,(m+10^{11}\, m^{2}+10^{27}\, m^{4})^{-0.36} \\  
    & && + 1.3\cdot 10^{-16} \, (m+10^{6}\, m^{2})^{-0.85}
\end{align}
is used. The mass influx in a mass range between the masses $m_1$ and $m_2$ can be calculated by integrating over the flux $F(m)$ and multiplying with Earth's surface $S_\mathrm{E} = 4\pi\cdot (6.471\cdot 10^3\,m)^2$ and the gravity enhancement factor $G=1.445$ \citep[see][]{Art_Drolshagen_2017}. Here, an incident atmospheric altitude of 100\,km is used. Additionally, the number of seconds in a year $T=3.1536\cdot 10^7$ has to be multiplied to yield the annual mass influx, then given as 
\begin{equation}
    M_{\mathrm{Gr\ddot{u}n}}=S_\mathrm{E} \cdot G\cdot T \cdot \int^{m_2}_{m_1}F(m)\,\mathrm{d}m.
\end{equation}

In the mass range from $10^{-2}$ to $10^{6}$\,kg and $10^{6}$ to $10^{11}$\,kg, the power laws from \cite{Art_Brown_2002} and \cite{Tech_Stokes_2003}
\begin{align}
    N_{\mathrm{Brown}}(E) &{}= 3.7\, E^{-0.9} \\
    N_{\mathrm{Stokes}}(E) &{}= 2.4\, E^{-0.79}
\end{align}
are used, respectively. The latter one is obtained from \cite{Art_Drolshagen_2017}. $N(E)$ represents the cumulative number of meteoroids with a kinetic energy greater than $E$ impacting Earth every year, where $E$ is in units of kt TNT equivalent. With an average meteoroid velocity of 20\,km/s and 1\,kt TNT equivalent $= 4.184\cdot 10^{12}$\,J, the energy dependence can be transformed to a dependence of mass:
\begin{align}
    N_{\mathrm{Brown}}(m) &{}= 2.86\cdot 10^4\,m^{-0.9} \\
    N_{\mathrm{Stokes}}(m) &{}= 6.22\cdot 10^4\,m^{-0.79}
\end{align} 
with $m$ the mass in kg. So $N(m)$ represents the number of meteoroids of a mass greater than $m$ hitting Earth every year.

\begin{table}[!th]
\centering
\caption{Annual mass influx of meteoroids into Earth's atmosphere for each mass decade (values of Figure \ref{fig1}).}
\label{ATAB1}
\small{
\renewcommand{\arraystretch}{1.2}
\begin{tabular}{cr@{}l | cr@{}l}
\toprule
log $m$ (kg) & \multicolumn{2}{c|}{$M$ (t/yr)} & log $m$ (kg) & \multicolumn{2}{c}{$M$ (t/yr)} \\
\midrule
-21 & 0&.02 & -5 & 384& \\
-20 & 0&.03 & -4 & 184& \\
-19 & 0&.04 & -3 & 86& \\ 
-18 & 0&.09 & -2 & 42& \\
-17 & 0&.26 & -1 & 53& \\
-16 & 1&.0 & 0  & 67& \\
-15 & 4&.3 & 1  & 84& \\
-14 & 18& & 2  & 106& \\
-13 & 85& & 3  & 133& \\
-12 & 367& & 4  & 168& \\
-11 & 1,125& & 5  & 211& \\
-10 & 2,193& & 6  & (264&) \\
-9  & 2,671& & 7  & (429&) \\
-8  & 2,213& & 8  & (696&) \\
-7  & 1,410& & 9  & (1,129&) \\
-6  & 768& & 10 & (1,831&) \\
\bottomrule
\end{tabular}
}
\renewcommand{\arraystretch}{1}
\caption*{The mass influx $M$ is given for the interval of the object mass $\mathrm{log}\,m$ to $\mathrm{log}\,(m) + 1$. For example, 384\,t of meteoroids in the mass range from 10$^{-5}$ to 10$^{-4}$\,kg impact Earth every year. The values derived from \cite{Tech_Stokes_2003} are in parenthesis as they are not included in the annual mass influx in this study.}
\end{table}
To yield the annual mass influx from these power laws, further calculations are necessary \citep[e.\,g. compare with][Appendix A]{Art_Bland_1996}. The number of meteoroids impacting per year in a mass range from $m_1$ to $m_2$ can be expressed by
\begin{equation}
\label{eqA5}
    N(m_1)-N(m_2) = -\int_{m_1}^{m_2}\frac{\mathrm{d}N(m)}{\mathrm{d}m}\,\mathrm{d}m.
\end{equation}
To yield the annual mass influx $M$ in the respective mass range, the mass has to be incorporated in the integral by multiplication:
\begin{equation}
\label{eqA6}
    M = -\int_{m_1}^{m_2}m\frac{\mathrm{d}N(m)}{\mathrm{d}m}\,\mathrm{d}m.
\end{equation}
This way, we yield the annual mass influx for both models in the mass range from $m_1$ to $m_2$
\begin{align}
    M_{\mathrm{Brown}} &{}= \int_{m_1}^{m_2} 2.57\cdot 10^4\,m^{-0.9}\,\mathrm{d}m \\
    M_{\mathrm{Stokes}} &{}= \int_{m_1}^{m_2} 4.9\cdot 10^3\,m^{-0.79}\,\mathrm{d}m.
\end{align}

Taking the respective valid mass ranges of each model (given above and in Figure \ref{fig1}), the integration yields $M_{\mathrm{Gr\ddot{u}n}} = 11,509$\,t/yr, $M_{\mathrm{Brown}} = 863$\,t/yr, and $M_{\mathrm{Stokes}} = 4,349$\,t/yr. The values for each mass decade, depicted in Figure \ref{fig1}, are given in Table \ref{ATAB1}. Remember that only the mass values from $M_{\mathrm{Gr\ddot{u}n}}$ and $M_{\mathrm{Brown}}$ (so masses in a range from $10^{-21}$ to $10^{6}$\,kg) are counted to the annual mass influx used in this paper.

\section{}\label{APP2}
The elemental mass abundances of meteorites are given in Table \ref{ATAB2}. Compositions are derived from the following sources: Chondrites, namely ordinary chondrites (H, L, and LL), carbonaceous chondrites (CH, CI, CK, CO, CR), Kakangari and Rumurutiites chondrites (K and R), and enstatite chondrites (EH and EL) from \citet[Tables 16.10 and 16.11]{Book_PlScCo_1998}; most achondrites, namely Acapulcoites (Acap), Angrites (Angr), Aubrites (Aubr), Brachinites (Brac), Diogenites (Diog), Eucrites (Eucr), Howardites (How), Lodranites (Lodr), Shergottites (Sher), Nakhlatites (Nak), Chassignites (Chas), Ureilites (Ur), and Winonaites (Wino) from \citet[Tables 16.11, 16.17, and 16.18]{Book_PlScCo_1998} and \citet[Tables 6, 8(4), 19, 21, 22, 26, 34, 35, and 40]{inCol_Mittlefehldt_2001}; lunar achondrites, so called Lunaites (Luna) from \cite{Art_Demidova_2007}; stony irons, namely Mesosiderites (Meso) and Pallasites (Pal) from \citet[Tables 13 (main group), 14 (main group), 15 (main group), 16 (all except Eagle Station), 17 (all except Eagle Station), 45(1), 46 (all except the last three)]{inCol_Mittlefehldt_2001} and \citet[Tables II-5 and II-7(Ni)]{Book_Meteorites_1974}; and for Irons (IAB to IVB) from \citet[Tables 3 and 8 (1 \& 9)]{inCol_Mittlefehldt_2001} and \citet[Table II-5]{Book_Meteorites_1974}. In a few cases, some abundances were estimated (mainly oxygen) considering similar meteorite subgroups to complement the data. The total mass abundance of all meteorites is shown in the last column. It is the product of the meteorite abundance with the respective elemental composition normalized to 100\,\% of the mass. Thus, it represents the overall elemental composition of meteorites found on Earth. This is used as the average elemental composition of meteorites given in the last column of Table \ref{TAB1}.

\begin{table*}[!t]
\centering
\caption{Elemental compositions of meteorite groups.}
\scriptsize{
\resizebox{2\columnwidth}{!}{
\begin{tabular}{@{}lll|lllllllllllllll@{}}
\toprule
\textbf{Z}           & \textbf{El.}           & \textbf{Unit} & \textbf{H} & \textbf{L} & \textbf{LL} & \textbf{CH} & \textbf{CI} & \textbf{CK} & \textbf{CM} & \textbf{CO} & \textbf{CR} & \textbf{CV} & \textbf{EH} & \textbf{EL} & \textbf{K}  & \textbf{R}  & \textbf{Acap} \\ \midrule
\multicolumn{2}{l}{Fraction $\rightarrow$} & wt\%          & 42.27      & 37.72      & 6.36        & 0.08        & 0.04        & 0.54        & 1.18        & 0.62        & 0.57        & 0.36        & 0.93        & 0.28        & 0.13        & 0.85        & 0.07          \\ \midrule
1                    & H                      & $\mu$g/g      &            &            &             &             & 20200       &             & 14000       & 700         &             & 2800        &             &             &             &             &               \\
3                    & Li                     & $\mu$g/g      & 1.7        & 1.85       & 1.8         &             & 1.5         & 1.4         & 1.5         & 1.8         &             & 1.7         & 1.9         & 0.7         &             &             &               \\
4                    & Be                     & ng/g          & 30         & 40         & 45          &             & 25          &             & 40          &             &             & 50          & 21          &             &             &             &               \\
5                    & B                      & ng/g          & 400        & 400        & 700         &             & 870         &             & 480         &             &             & 300         & 1000        &             &             &             &               \\
6                    & C                      & $\mu$g/g      & 2100       & 2500       & 3100        & 7800        & 34500       & 2200        & 22000       & 4400        & 20000       & 5300        & 3900        & 4300        &             & 580         &               \\
\bottomrule
\end{tabular}
}
}
\label{ATAB2}
\end{table*}

\begin{table*}[!t]
\caption*{{Table \ref{ATAB2} \textit{(continued)}}}
\centering
\scriptsize{
\resizebox{2\columnwidth}{!}{\begin{tabular}{@{}lll|lllllllllllllll@{}}
\toprule
\textbf{Z}           & \textbf{El.}           & \textbf{Unit} & \textbf{Angr} & \textbf{Aubr} & \textbf{Brac} & \textbf{Diog} & \textbf{Eucr} & \textbf{How}  & \textbf{Lodr} & \textbf{Luna} & \textbf{Sher} & \textbf{Nak} & \textbf{Chas} & \textbf{Ur} & \textbf{Wino} & \textbf{Meso} & \textbf{Pal} \\ \midrule
\multicolumn{2}{l}{Fraction $\rightarrow$} & wt\%          & 0.02          & 0.28          & 0.04          & 0.57    & 1.22          & 0.57          & 0.09          & 0.11          & 0.03          & 0.03         & 0.03          & 0.56        & 0.07          & 0.29          & 0.22         \\ \midrule
1                    & H                      & $\mu$g/g      &               &               &               &               &               &               &               &               &               &              &               &             &               &               &              \\
3                    & Li                     & $\mu$g/g      &               &               &               &               & 5.4           & 3.5           &               &               & 3.7           & 3.9          & 1.4           &             &               &               &              \\
4                    & Be                     & ng/g          &               &               &               &               &               &               &               &               &               &              &               &             &               &               &              \\
5                    & B                      & ng/g          &               &               &               &               &               &               &               &               &               &              &               &             &               &               &              \\
\bottomrule
\end{tabular}
}
}
\end{table*}

\begin{table*}[!t]
\caption*{{Table \ref{ATAB2} \textit{(continued)}}}
\centering
\scriptsize{
\resizebox{1.92\columnwidth}{!}{\begin{tabular}{@{}lll|lllllllllllll|l@{}}
\toprule
\textbf{Z}           & \textbf{El.}           & \textbf{Unit} & \textbf{IAB} & \textbf{IC} & \textbf{IIAB} & \textbf{IIC} & \textbf{IID} & \textbf{IIE} & \textbf{IIF} & \textbf{IIIAB} & \textbf{IIICD} & \textbf{IIIE} & \textbf{IIIF} & \textbf{IVA} & \textbf{IVB} & \textbf{Mean} \\ \midrule
\multicolumn{2}{l}{Fraction $\rightarrow$} & wt\%          & 0.76         & 0.06        & 0.60          & 0.05         & 0.09         & 0.11         & 0.03         & 1.34           & 0.24           & 0.08          & 0.04          & 0.37         & 0.08         & 100.00                 \\ \midrule
1                    & H                      & $\mu$g/g      &              &             &               &              &              &              &              &                &                &               &               &              &              & 185.6                  \\
3                    & Li                     & $\mu$g/g      &              &             &               &              &              &              &              &                &                &               &               &              &              & 1.66                   \\
4                    & Be                     & ng/g          &              &             &               &              &              &              &              &                &                &               &               &              &              & 31                     \\
5                    & B                      & ng/g          &              &             &               &              &              &              &              &                &                &               &               &              &              & 377                    \\
\bottomrule
\end{tabular}
}
}
\end{table*}

\cleardoublepage
\newpage

\onecolumn

\section{}\label{APP3}
Table \ref{ATAB3} provides an overview over the rocket launches of 2019 and the respective reentering core stages. \nopagebreak
\begin{table*}[!h]
\caption{Rocket launches of 2019 with the mass and velocity of reentering core stages. Only stages with a significant velocity are considered.}
\label{ATAB3}
\centering
\renewcommand{\arraystretch}{1}
\scriptsize{
\begin{tabular}{@{}lp{1.3cm}p{1.2cm} p{1.3cm} p{1.2cm} p{1.7cm} p{5.3cm}@{}}
\toprule
Launch vehicle           & Launches (2019)    & Stage number   & Stage mass (t)   & 2019 mass (t) & Reentry speed (km/s) & Sources \\ \midrule
Soyuz 2.1a,b/Soyuz-FG      & 7                  & 2                     & 6.55                  & 45.8           & 4.3                     &   \cite{Misc_youtube_soyuz, Tech_RS_Soyuz}     \\ \midrule
\multirow{2}{3.3cm}{Soyuz 2.1a,b/Soyuz-FG with Fregat upper stage}     & \multirow{2}{*}{9}                  & 2                     & 6.55                  & 58.9           & 3.9                     &   \multirow{2}{4cm}{\cite{Misc_youtube_soyuz_fregat, Tech_RS_Soyuz} }    \\ 
     &                   & 3                     & 2.36                  & 21.2           & 7.0                     &        \\ \midrule
Soyuz 2.1v Volga     & 2                  & 2                     & 2.36                  & 4.7           &                      &   \cite{Misc_soyuz2.1v_gunter, Tech_RS_Soyuz}     \\ \midrule
Falcon 9\textsuperscript{a}                 & 11                  &                      &                       & 0             &                         &         \\ \midrule
\multirow{2}{*}{Proton}          & \multirow{2}{*}{5}                    & 2                     & 11.0                    & 55.0            & 4.6                     &   \multirow{2}{4cm}{\cite{Tech_RS_proton} }     \\
         &                    & 3                     & 3.5                   & 17.5            & 7.3                     &         \\
\bottomrule
\end{tabular}
}
\renewcommand{\arraystretch}{1}
\caption*{The number of launches of each launch vehicle is retrieved from \cite{Misc_RS_LaunchReport}, for the chinese rockets (CZ-X) also from \cite{Misc_cz2to4_gunter}. Numbers in parenthesis in column 2 depict the number of failures. The stage numbering in column 3 considers booster stages as first stages. The stages' masses are retrieved mostly from data sheets and information available online. The reentry velocity is calculated from the velocity and altitude at stage separation (if data is available). Small rockets and suborbital flights are given but neglected in the calculations as the reentering mass is comparably low.  \newline
\textsuperscript{a} First stage (for Falcon Heavy also the second stages) perform a controlled landing, thus do not ablate in the atmosphere. \newline
\textsuperscript{b} Given that the rocket is configured with boosters, otherwise this is stage 1.}
\end{table*}







\cleardoublepage
\newpage
\bibliographystyle{model2-names}\biboptions{authoryear}
\bibliography{bib.bib}{}
 

\end{document}